\documentclass[twocolumn]{revtex4}

\usepackage{graphicx}

\newcommand{\beq}{\begin{equation}}
\newcommand{\enq}{\end{equation}}
\newcommand{\bea}{\begin{eqnarray}}
\newcommand{\ena}{\end{eqnarray}}

\newcommand{\rr}{{\bf r}}

\newcommand{\aos}{a_{\mathrm{osc}}}

\begin{document}
\title{Collective modes and the broken symmetry of a 
rotating attractive Bose gas in an anharmonic trap}
\author{A. Collin}
\affiliation{Helsinki Institute of Physics, PL 64, FIN-00014
Helsingin yliopisto, Finland}
\date{\today}

\begin{abstract}
We study the rotational properties of an attractively interacting Bose 
gas in a quadratic + quartic potential. The low-lying modes of both
rotational ground state configurations, namely the vortex and 
the center of mass rotating states, are solved. The vortex excitation 
spectrum is positive for weak interactions but the lowest modes decrease 
rapidly to negative values when the interactions become stronger.
The broken rotational symmetry involved in the center of mass rotating 
state induces the appearance of an extra zero-energy mode in the 
Bogoliubov spectrum. The excitations of the center of mass rotational 
state also demonstrate the coupling between the center of mass 
and relative motions.
\end{abstract}

\maketitle

\section{Introduction}

The appearance of quantized vortices offers a strong evidence of 
superfluidity in a rotating Bose-Einstein condensed system. 
A classical example is the quantized circulation in superfluid $^4$He. 
More recently, the existence of these novel quantum states has 
also been demonstrated in dilute atomic gases
\cite{Pethicksmith2001,Pitastring2003}. It is found that
above a certain critical rotation frequency, atomic vapor 
with {\it repulsive} interactions responds to rotation by forming a 
singly quantized vortex \cite{Matthews1999,Madison2000}. 
As the rotation rate increases, more vortices appear and
vortex lattices consisting hundreds of vortices may be detected
\cite{Abo2000,Engels2003}. In addition to atomic boson gases,
the experimental detection of vortex lattices in both 
sides of Feshbach resonance in quantum degenerate Fermi 
gas has just been reported \cite{Zwierlein2005}.

The response of a Bose condensed gas to an external rotation 
is expected to be different in the  case  of {\it attractive} interactions. 
In a harmonic-oscillator potential, it is not the vortex state
which is energetically favorable. Instead, the least-energy  
configuration for an attractive condensate with non-zero angular momentum 
is a state with a rotating center of mass (c.m.) 
\cite{Wilkin1998,Mottelson1999,Pethick2000}. However, this state is 
thermodynamically unstable, for the  critical rotation 
frequency to excite the c.m. mode is equal to 
the trap frequency. The condensate set on rotation thus experiences 
the centrifugal force being larger than the confining force of the trap and
eventually escapes the trap.

Although a rotating attractive condensate is not stable in 
a harmonic potential, it can be stabilized by using a steeper 
confinement. A possible scheme is an anharmonic trap configuration 
in which a small quartic potential is superposed onto the harmonic trap.
This potential is experimentally realistic; it has already been implemented 
to study the fast rotation of repulsive $^{87}$Rb 
condensate~\cite{Bretin2004}. On the attractive side it is expected 
that the phase-space consists of states of broken cylindrical symmetry and, 
for very weak interactions, multiply quantized vortices 
\cite{Us2004,Kavo2004,Ghosh2004,Us2005}. The states of broken 
cylindrical symmetry
involve a rotating center of mass and are closely connected with
the c.m. rotating state in a harmonic trap. Even though the correspondence    
is never exact, in this paper we use the term c.m. rotating state 
for all the states with a broken cylindrical symmetry.

The rich collection of rotational states of an attractive condensate 
in an anharmonic trap is worth further studies. An especially interesting 
effect is the spontaneously broken cylindrical symmetry in the case of c.m. 
rotating state. In addition, the investigation of dynamical 
and thermodynamical processes is a necessary extension to the work 
performed so far. This includes the study of the lack of superfluidity
in attractive condensates \cite{Leggett2001} indicating the fact
that, if the external rotation is stopped, the condensate 
relaxes to the non-rotating ground state, independently 
of the initial rotational state. Recently, these topics have
been studied in one dimension where even exact results have been obtained   
\cite{Kana2003,Kartsev2003,Kavo2004b,Kana2005}.

In the present work, we study the elementary excitations of rotating 
attractive Bose condensed gas confined in an anharmonic quartic + quadratic 
trap potential. We present our system and the relevant formalism 
in Sec.~\ref{sec:eqs}. The nature of the ground state excitation spectrum is
discussed in Sec.~\ref{sec:gsp}, and the results are summarized in 
Sec.~\ref{sec:con}. 
 
\section{The model}
\label{sec:eqs}

We assume a zero-temperature Bose gas with negative
$s$-wave interaction. The anharmonic confining potential has 
the following form
\beq
\label{anharmpotential}
   V(r,\theta,z)=\frac12m\left[\omega^2 \left(r^2 +
   \lambda \frac{r^4}{\aos^2}\right)
   +\omega_z^2 z^2\right],
\enq 
which consists of the harmonic oscillator potential plus
the radial quartic term described by the dimensionless parameter $\lambda$. 
The oscillator length is defined as $\aos=(\hbar/m\omega)^{1/2}$.

Following the argumentation of Ref.~\cite{Us2005}, by ignoring
the quantum fluctuations of the possible c.m. motion, the 
ground state behavior of a rotated attractive Bose condensate 
can be approximated by the ordinary single-component 
Gross-Pitaevskii (GP) equation in the frame of reference 
rotating with frequency ${\bf \Omega}$. 
The GP equation for a stationary state is 
\beq
\label{GP}
\left[H_0-
{\bf \Omega}\cdot\hat{L}+U_0|\Psi\left(\rr\right)|^2\right]
\Psi\left(\rr\right)
=\mu\Psi\left(\rr\right),
\enq
where $H_0=-\frac{\hbar^2}{2m}\nabla^2+V\left(\rr\right)$ is
the Hamiltonian for a single atom in the nonrotating trap. The interaction
constant $U_0$ is proportional to the scattering length by the relation 
$U_0=4\pi\hbar^2a/m$.
$\hat{L}=-i\hbar\left(\rr\times\nabla\right)$ is
the angular momentum operator and $\mu$ is the chemical potential.
The condensate wave function $\Psi$ is normalized to the number
of atoms $N$ which is fixed here to $N=1000$. We also fix
the mass $m$ to the mass of atomic $^{7}$Li. For the angular frequencies
of the harmonic oscillation we set the values $\omega=2\pi\times 30$ Hz and 
$\omega_z=2\pi\times 180$ Hz. Here, ${\bf \Omega}$ is always pointing
to the $z$-direction, which, together with the strong confinement in
this direction, makes the problem effectively two-dimensional.  

The phase-diagram of the ground state solutions of Eq.~(\ref{GP}) is studied
in \cite{Us2005}. Here, we extend our earlier work and study the 
spectrum of normal modes of the solutions of Eq. (\ref{GP}).
These are obtained by solving the coupled Bogoliubov-de Gennes equations 
for the quasiparticles $u_j$ and $v_j$,
\bea
\label{Bdg}
Ku_j\left(\rr\right)
+U_0\Psi^2\left(\rr\right)v_j\left(\rr\right) & = & E_ju_j
\left(\rr\right),
\nonumber\\
K^*v_j\left(\rr\right)
+U_0\Psi^{*2}\left(\rr\right)u_j\left(\rr\right) & = & -E_jv_j
\left(\rr\right),
\ena 
where $K=H_0-{\bf \Omega}\cdot\hat{L}+2U_0|\Psi\left(\rr\right)|^2 -\mu$
and $E_j$ is the energy of the $j$th eigenmode. 

The calculation of the spectrum of the eigenmodes in our numerical procedure
consists of first solving the lowest energy configuration of Eq. (\ref{GP})
in a two-dimensional Cartesian grid by propagating the wave function
in imaginary time. The second step is to transfer the obtained wave function
and chemical potential to Eqs. (\ref{Bdg}) and numerically diagonalize
these in order to solve $u_j$, $v_j$ and $E_j$.

\section{ground state excitations}
\label{sec:gsp}

\begin{figure}
\includegraphics[width=8.0cm]{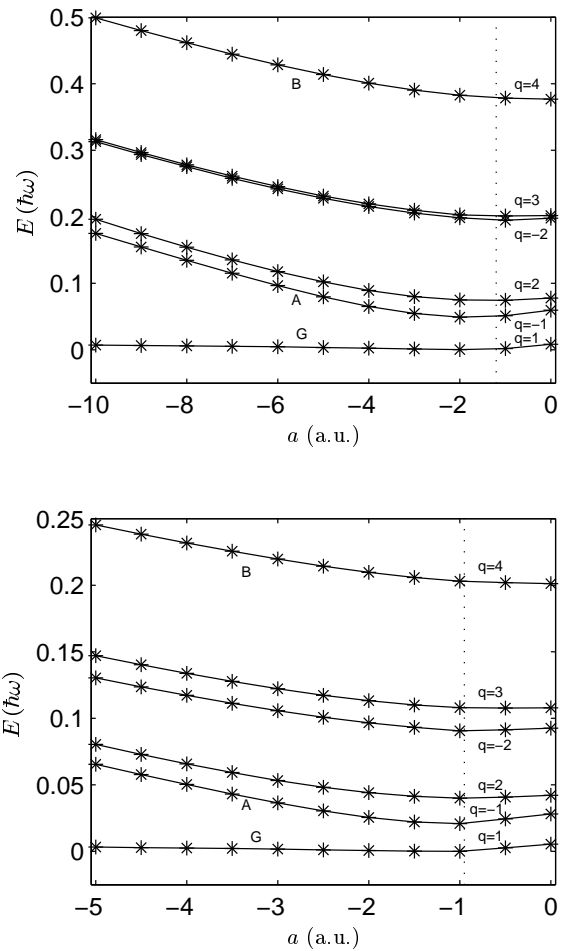}
\caption{The lowest quasiparticle eigenenergies for ground
state configurations calculated for different values of the 
scattering length $a$. In the upper graph ($\lambda=0.15,\Omega/\omega=1.35$),
and ($\lambda=0.05,\Omega/\omega=1.15$) in the
lower graph. The dotted vertical line represents the interpolated 
value of the critical scattering length $a_c$ separating the 
$m=2$ vortex ground state (right) and the c.m. ground state (left).}
\label{evs015}
\end{figure}

We analyze the numerical solutions of Eq.~(\ref{Bdg}) by fixing 
the pair of parameters ($\lambda$,~$\Omega/\omega$) and varying the 
interaction strength $U_0$. We first choose 
($\lambda=0.15$,~$\Omega/\omega=1.35$). For these parameters 
the phase-space was analyzed in Ref.~\cite{Us2005}. 
In addition, we choose another pair ($\lambda=0.05$,~$\Omega/\omega=1.15$)
to be studied, for these values are closer to the pure harmonic 
oscillator. The low-energy spectra are shown in Fig. \ref{evs015} 
where we have varied the scattering length $a$ to cover the 
phase-space portions of vortex and c.m. states. In both cases, the 
ground state on the vortex side is doubly quantized. A vortex is 
only stable for weak interactions, and the increase
in the magnitude of interaction strength leads to a phase transition
at certain critical $a=a_c$. For these relatively strong interactions,
the ground state is the one with broken rotational symmetry. For
very strong interactions (or high $N$) the system becomes unstable.

\subsection{Vortex ground state}
 
On the vortex side, the ground state as well as the excitations are 
cylindrically symmetric. Consequently, the Bogoliubov amplitudes 
can be written as $u_q\propto e^{i\left(q+m\right)\theta}$  with  
$v_q\propto e^{i\left(q-m\right)\theta}$. Here $m$ (not to be confused 
with the mass) is the angular momentum quantum number of the condensate, 
which, in the case of a doubly quantized vortex is $m=2$. 
The excitation winding number $q$ is the angular momentum 
relative to the condensate. Energy spectra in Fig. \ref{evs015} show
that the lowest excitations are the modes with $q=1$ and $q=-1$. 
Especially, the energy eigenvalue of $q=1$ mode is very low. 
Increasing $|a|$ decreases the energy of the mode quickly to 
negative values. The crossover determines the critical value $a_c$
below which the vortex state is not stable. At $a\approx a_c$
the mode energy is real within the numerical precision. However,
in the unstable region the imaginary part of the negative mode
rapidly increases as a function of the interaction strength.
This indicates the dynamical instability of the vortex state.
We have checked this by performing time evolution GP simulations. 
Indeed, we have seen the breakup of the vortex wave 
function when $|a|$ is strong enough whereas for $|a|<a_c$ the 
vortex state is stable in time.  

\subsection{C.m. ground state}

\begin{figure}
\includegraphics[width=8.0cm]{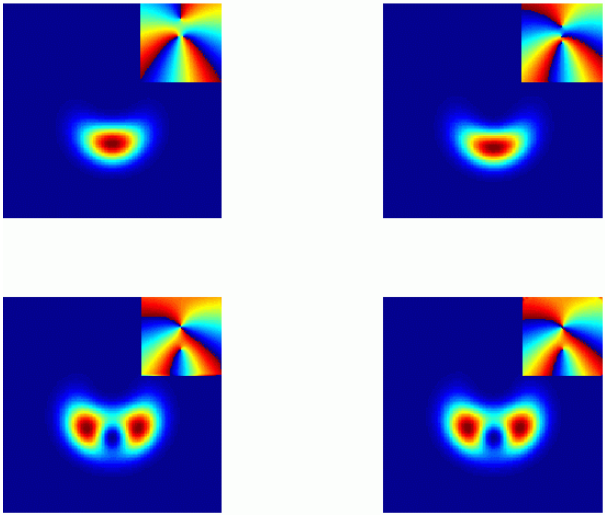}
\caption{Density plots of different wave functions related to
the c.m. state (upper left graph). The relevant parameters
are $\lambda=0.15$, $\Omega/\omega=1.35$ and $a=-10.0$ a.u. The upper 
right graph is the Goldstone mode $\Phi_\theta$ as defined in   
Eq. (\ref{gse}). The part of $\Phi_\theta$ orthogonal to  
the c.m. rotational wave function is denoted as $\chi$ and plotted in the
lower left graph. In the lower right graph we plot the
zero-energy Bogoliubov solution $u_G$. The insets show 
the phase profiles of the wave-functions.}
\label{gsm}
\end{figure}

On the left hand side of the Fig. \ref{evs015} the excitations are 
calculated for the ground state solution which is the c.m. rotational state. 
The ground state stability is ensured by the fact that the excitation 
energies are real and non-negative. We have also propagated
the GP ground state solutions in real-time and we have not seen 
any sign of a dynamical instability.    

As seen in Fig. \ref{evs015}, there is an extra zero-energy 
mode on the c.m. side of the phase-space (denoted by G). 
Due to Goldstone's theorem \cite{Goldstone}, the spontaneously broken 
symmetry implies the existence of a zero energy mode.
Generally, the mean-field approximation breaks the global U(1)
symmetry of the system 
\footnote{In the context of Bogoliubov amplitudes 
this mode is of the form $u=\alpha\Psi$ and $v=-\alpha\Psi$ 
where $\alpha$ is a complex constant. The numerical procedure gives 
this trivial solution, but we have not included it into the spectrum 
in Fig. \ref{evs015}.}. However, in the case of c.m. rotational state, 
the cylindrical O(2) symmetry is also spontaneously broken. 
For a mean-field wave function $\phi_c$ with a particular 
broken symmetry $\varphi_0$, the Goldstone mode is 
$\frac{\partial\phi_c\left(\rr,\varphi_0\right)}{\partial\varphi_0}$
\cite{Negele}. In this case, the Goldstone mode corresponding to 
the rotational asymmetry is then
\beq
\label{gse}
\Phi_\theta\left(\rr\right)=
\frac{\partial\Psi\left(\rr\right)}{\partial\theta}. 
\enq
In Fig. \ref{gsm} we show an example of a c.m. rotational wave function and 
the corresponding Goldstone mode $\Phi_\theta\left(\rr\right)$.
The density profile of $\Phi_\theta\left(\rr\right)$ is slightly 
more elongated, but otherwise it is rather similar to 
$\left|\Psi\left(\rr\right)\right|^2$. The structure of 
$\Phi_\theta\left(\rr\right)$ may be confusing at first 
sight, but there is an extra twist in the problem. Namely, if more than 
one symmetry is spontaneously broken, the corresponding Goldstone 
modes are not necessarily orthogonal. This means that the
mode in Eq.~(\ref{gse}) is generally a superposition of the c.m. 
rotational wave function $\Psi$ and a wave function 
$\chi$ orthogonal to $\Psi$. In this context, we write
\beq
\Phi_\theta\left(\rr\right)=\beta\Psi\left(\rr\right)+\chi\left(\rr\right),
\enq 
where $\beta$ is a complex constant. Because of orthogonality, 
both $\chi$ and $\beta$ can be easily solved. In the example
of Fig. \ref{gsm}, we present the function $\chi$ in the left lower graph. 
The extra zero-energy Bogoliubov solution for the same wave 
function can also be written as a superposition 
of the two orthogonal modes. Also, the mode amplitude $u_G$ has a 
component orthogonal to the c.m. rotational wave function. This is plotted 
in the right lower graph in Fig. \ref{gsm}. The identical structure 
(within numerical precision) with the function $\chi$ is clearly
visible. 

The differential operation performed on the
condensate wave function in Eq. (\ref{gse}) is just $i\hat L$.
Although the function $\chi$ is generally small compared to 
$\beta\Psi$ it is clear that $\Psi$ is not an angular momentum 
eigenstate. On the other hand,  the true many-body c.m. 
rotational state in a harmonic trap \cite{Wilkin1998} is an 
eigenstate of $\hat L$. The contradiction is a side effect 
of the broken symmetry, which, we believe, in turn is a consequence of the 
used GP approximation. This is supported by the quasi-one-dimensional 
analog \cite{Kana2003,Kavo2004b,Kana2005}, where the symmetry is broken 
independent of the details of the trap potential which forms the torus.

\begin{figure}
\includegraphics[width=8.0cm]{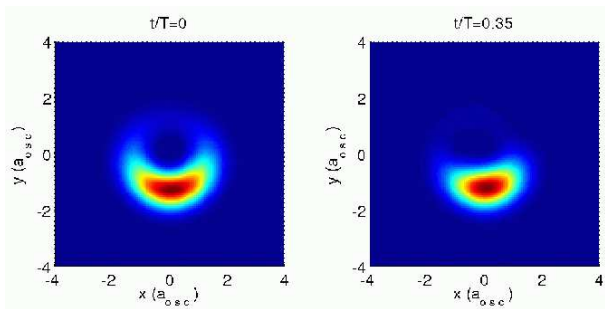}
\caption{A c.m. rotational ground state excited with the lowest 
true excitation (A) obtained from Bogoliubov-de Gennes equations. 
Here $\lambda=0.15$ and $a=-5.0$ a.u. and $T\approx 78~1/\omega$ 
is the period of oscillation.}
\label{bre}
\end{figure}

For actual, positive energy excitations the general behavior is
that the energy eigenvalues increase with the interaction strength.
However, interpretations of individual modes should be done with care. 
In one-dimensional torus, the lowest mode is a breathing mode 
\cite{Kana2005}. Generally, this is not the case in 2D because 
the anharmonic term in the trap potential couples the c.m. and 
relative motions. A change in the mean size of the cloud 
affects the effective potential seen by the c.m. Consequently,
a time dependent wave function corresponding to an excited mode 
is expected to show some c.m. motion also. For the 
lowest excitation (denoted by A in Fig. \ref{evs015}), 
the coupling can be seen in
Fig. \ref{bre} where we have assumed an excited wave function
of the form $\Psi\left(\rr,t\right)=
\Psi\left(\rr\right)+u_Ae^{-iE_At/\hbar}-v_A^*e^{iE_At/\hbar}$.
As a rough description of the oscillation, 
at $t/T=0$ the elongated wave function is on azimuthal motion. 
The c.m. motion then begins to decelerate and
excitational breathing begins to compress the cloud. The compression 
rate is highest at the classical turning point of the c.m. On the backward
c.m. motion the cloud retains the squeezed shape (see $t/T=0.35$ plot 
in Fig. \ref{bre}). When approaching the other turning 
point the c.m. motion slows down and the cloud begins to enlarge again.
The coupling between c.m. and shape oscillations is not 
characteristic only to this mode but also to all other modes 
we have investigated.  

The investigation of the mode profiles for $u_j$ and $v_j$ also 
reveals that the broken rotational symmetry is not as 
visible for the high energy modes as it is for the low-lying ones. 
In the computations we have performed, the mode B
corresponding to the $q=4$ mode on the vortex side carries only a 
slight asymmetry on the c.m. side whereas the lowest excitation
is highly localized.

\section{Conclusions}
\label{sec:con}
In summary, we have studied the Bogoliubov-de Gennes excitation spectrum
of a rotating attractively interacting Bose-Einstein condensate  
trapped in an anharmonic potential. The stability of the (multiply)
quantized vortex state is ensured by the positive energy of the
lowest excitations. However, the eigenmodes $q=1$ and $q=-1$
become unstable at relatively weak values of the scattering length.
These modes being unstable the true ground state becomes the
state with the rotating center of mass. For the c.m. state the
rotational symmetry is broken and the Bogoliubov spectrum reveals an
extra zero-energy eigenmode. This Goldstone mode is the rotation
mode of the c.m. state. Finally, by exciting the c.m. state with
lowest positive energy mode, we have illustrated the coupling
between the c.m. and relative motions in anharmonic potential.

\section{Acknowledgments}
Discussions with E. Lundh, M. Mackie, J.-P. Martikainen and 
K.-A. Suominen are gratefully acknowledged. This work is supported
by Magnus Ehrnrooth foundation and the Academy of 
Finland (Project No. 206108).

\end{document}